*Review*

# POLYMER-BASED SOLID-STATE ELECTROLYTES FOR LITHIUM SULFUR BATTERIES

**Praveen Balaji T[1,2], Soumyadip Choudhury[1], Ernesto E. Marinero[2*]**

[1] Rubber Technology Centre, Indian Institute of Technology Kharagpur, India – 721302
[2] School of Materials Engineering, Purdue University, West Lafayette, IN 47907
* Correspondence: eemarinero@purdue.edu

**Abstract:** Lithium-sulfur (Li-S) batteries offer substantial theoretical energy density gains over Li-ion batteries, a crucial factor for transportation electrification. In addition, sulfur is an earth-abundant, inexpensive material obtainable from multiple resources; thus, Li-S batteries are envisioned to provide environmentally sustainable solutions to the growing demand for energy storage. A critical roadblock to the realization of commercial Li-S batteries is the formation of polysulfides and their secondary reactions with liquid organic electrolytes, resulting in low coulombic efficiency for charging and fast self-discharge rates. The realization of solid-state electrolytes for Li-S batteries provides potential pathways to address the safety concerns of liquid electrolytes and inhibit the formation of polysulfides and/or prevent their diffusion into the anode electrode. However, current solid-state electrolytes are limited by low ionic conductivity, inadequate electrode interfacial compatibility, and restricted electrochemical windows. This review discusses the status of polymer-based electrolytes for Li-S batteries, and outlines current methods for their fabrication, their transport characteristics and ongoing research aimed at overcoming material properties hindering the development of all-solid-state Li-S batteries.

## 1. Introduction

Transportation electrification, advanced portable electronics, and grid storage have increased demand for higher energy-density secondary batteries. Furthermore, for a sustainable planet, it is imperative that battery systems (materials, fabrication, recycling) minimize global emissions to mitigate irreversible climate warming[1,2]. Lithium-ion batteries (LIBs) are currently the predominant energy storage devices available in the market, and electrode improvements enable reaching their theoretical gravimetric and volumetric energy densities, 350 $Wh.kg^{-1}$ and 700 $Wh.L^{-1}$, respectively[3]. The Lithium-sulfur (Li-S) battery is a potential candidate for higher energy density applications, with a theoretical energy density of 2600 $Wh.kg^{-1}$ and a specific capacity of 1675 $mAh.g^{-1}$[4,5]. The enhanced energy storage capacity of Li-S batteries derives from phase-transformation chemistry between $Li^+$ cations and sulfur ($S_8$) to form $Li_2S$[6]. This conversion reaction enables the accommodation of more $Li^+$ ions in the anode, resulting in higher energy storage capacity, and the utilization of sulfur, an earth-abundant, low-cost cathode material, will lower energy storage costs and an environmentally-friendy solution for beyond-Li ion batteries[7].

A Li-S battery comprises a cathode composed of elemental sulfur, a lithium metal anode, and an electrolyte[8]. Their operation involves a reversible redox reaction between Li+ cations and $S_8$, enabling the conversion of chemical energy into electrical energy [$S_8 + 6Li \leftrightarrow 8Li_2S$]. Throughout the discharge process, solid $S_8$ reactions with Li result in two distinct forms of lithium polysulfides (LPS): long-chain polysulfides, $Li_2Sn$ (where $4 < n \leq 8$), and short-chain polysulfides (where $2 < n \leq 4$). The polysulfides are further converted into insoluble compounds, specifically $Li_2S_2$ and $Li_2S$. In the context of a liquid electrolyte, both long-chain polysulfides and short-chain polysulfides are soluble, leading to secondary reactions and the loss of active sulfur, resulting in reduced efficiency and lifespan of the battery. This phenomenon is known as the shuttle effect[8–10]. The generation and dissociation of polysulfides are essential to forming $Li_2S$ and, thus, the operation of Li-S batteries. Dissolution of polysulfides provides sulfur atoms at the conductive carbon surface, driving the electrochemical reaction for battery operation. Polysulfide dissolution in liquid electrolytes is a required mechanism for Li-S battery operation. However, secondary reactions between polysulfides and electrolyte/electrode materials cause the shuttle effect. Polysulfides in solution can migrate to the anode and react with lithium metal, resulting in the formation of insoluble lithium sulfides that accumulate on the anode's surface. This process can result in creating an unstable solid electrolyte interphase (SEI) layer, reducing battery capacity, performance and coulombic efficiency and potentially leading to safety concerns on account of dendrite formation[11,12]. While the generation and dissolution of polysulfides are essential for battery functionality, effectively controlling their movement and reactivity is vital for attaining long-lasting stability and optimal performance. The organic solvents in liquid electrolytes have low flash points, high volatility, and flammability and can be readily ignited when subjected to abusive conditions such as overheating and overcharging, potentially resulting in catastrophic explosions[13]. Additionally, the insulating nature of sulfur and the sluggish redox reaction kinetics yield suboptimal electrochemical performance and inadequate cyclability[14]. Furthermore, electrode and electrolyte materials become depleted, and dendritic growth may occur, all posing significant limitations for Li-S batteries [19-21].

Two solutions have been advanced to address the shuttle effect in liquid electrolytes: a) incorporation of functional materials into either the cathode or the electrolyte[15]. Incorporating functional materials can impede the shuttle effect by spatially constraining and chemically tethering LPS, using catalysts to expedite reaction kinetics, and obstructing the diffusion of soluble polysulfides. Functional additives such as lithium nitrate ($LiNO_3$) react with polysulfides to form stable products, thus inhibiting their movement or diffusion[16]. Metal oxides can absorb polysulfides and limit their movement; b) deploying an interlayer between the cathode and the separator to mitigate the diffusion of polysulfides. The interlayer functions as a protective barrier preventing polysulfide dissolution and diffusion from the cathode to the anode. This process involves anchoring and converting soluble polysulfides into less mobile forms, impeding their movement and allowing Li+ ions to pass through easily. These measures merely mitigate rather than fully resolve the shuttle effect[17]. Several strategies have been developed to address these limitations, and the utilization of polymer-based electrolytes has emerged as a promising solution. These electrolytes serve a dual function of transporting Li+ ions and separating the cathode from the anode. The transition from liquid electrolytes to solid-state electrolytes is expected to enhance safety, increment energy density, and ultimately provide a solution to the shuttle effect[18]. Figures 1a and 1b provide schematic drawings of Li-S batteries employing liquid and polymer-based electrolytes, respectively.

Candidate materials for solid-state electrolytes (SSEs) include solid polymer electrolytes (SPE), composite polymer electrolytes (CPE), and inorganic solid electrolytes (ISE). To be effective electrolytes, SSEs should provide: 1) a high ion transference number[19]. 2) enhanced lithium-ion conductivity[20]. 3) a stable electrochemical window (between 0 and 5 V)[21]. 4) adequate mechanical flexibility that guarantees mechanical robustness[22]. 5) effective electrical insulation[23]. A comparison of these three types of electrolytes is provided in Figure 1c. Schematics of Li-S batteries with various electrolyte types: liquid, SPE, ISE and CPEs are shown in Figure 2c, which also provides radar plots of their current performance characteristics, which include polysulfide solubility, safety, interface compatibility, processability, mechanical strength, and ionic conductivity.

Polymer-based electrolytes provide potential significant advantages over liquid electrolytes for Li-S batteries. They are significantly less-flammable and mechanically robust, thereby enhancing the safety of Li-S batteries. Polymer electrolytes possess a higher degree of flexibility than inorganic counterparts, enabling improved electrode contact and accommodation of volume fluctuations during cycling[24]. Polymer electrolytes can hinder the diffusion of polysulfide intermediates, thereby reducing the shuttle effect[25]. Solid polymer electrolytes offer superior mechanical characteristics and can effectively inhibit the growth of lithium dendrites[26]. Their composition and microstructure and inherent transport and physical properties are readily modified, amongst others, by modifying their backbone composition, incorporating side chains, and employing polymer blends. Polymers can be readily combined with additional elements, such as fillers or ionic liquids to produce composite electrolytes to engineer and improve their transport and physico-chemical characteristics[27–30]. The excellent processability of polymers facilitates low-cost scalable production of thin-film electrolytes[31]. A comparison of different types of solid electrolytes is provided in Table 1.

This review focuses on polymer-based electrolyte materials for Li-S batteries and includes discussions on polymers, inorganic fillers, anion salts and processing methodologies. The most widely investigated polymer materials for electrolyte applications are polyethylene oxide (PEO)[32], polyvinylidene fluoride (PVDF)[33], polyacrylonitrile (PAN)[34], polyvinylidene fluoride hexafluoropropylene (PVDF-HFP)[35], polymethyl methacrylate (PMMA)[36], and its derivatives. Polymer electrolytes incorporating non-ion conducting ceramic fillers ($Al_2O_3$, $SiO_2$, and $TiO_2$), fast ion conducting ceramic particles, ionic liquid fillers, nanostructured fillers (nanowires, nanosheets), metal oxides, sulfides and hybrid fillers are also reviewed[37]. Anion salts added to polymers include Lithium bis(trifluoromethane)sulfonamide (LiTFSI)[38] and Lithium bis(fluorosulfonyl)imide (LiFSI)[39].

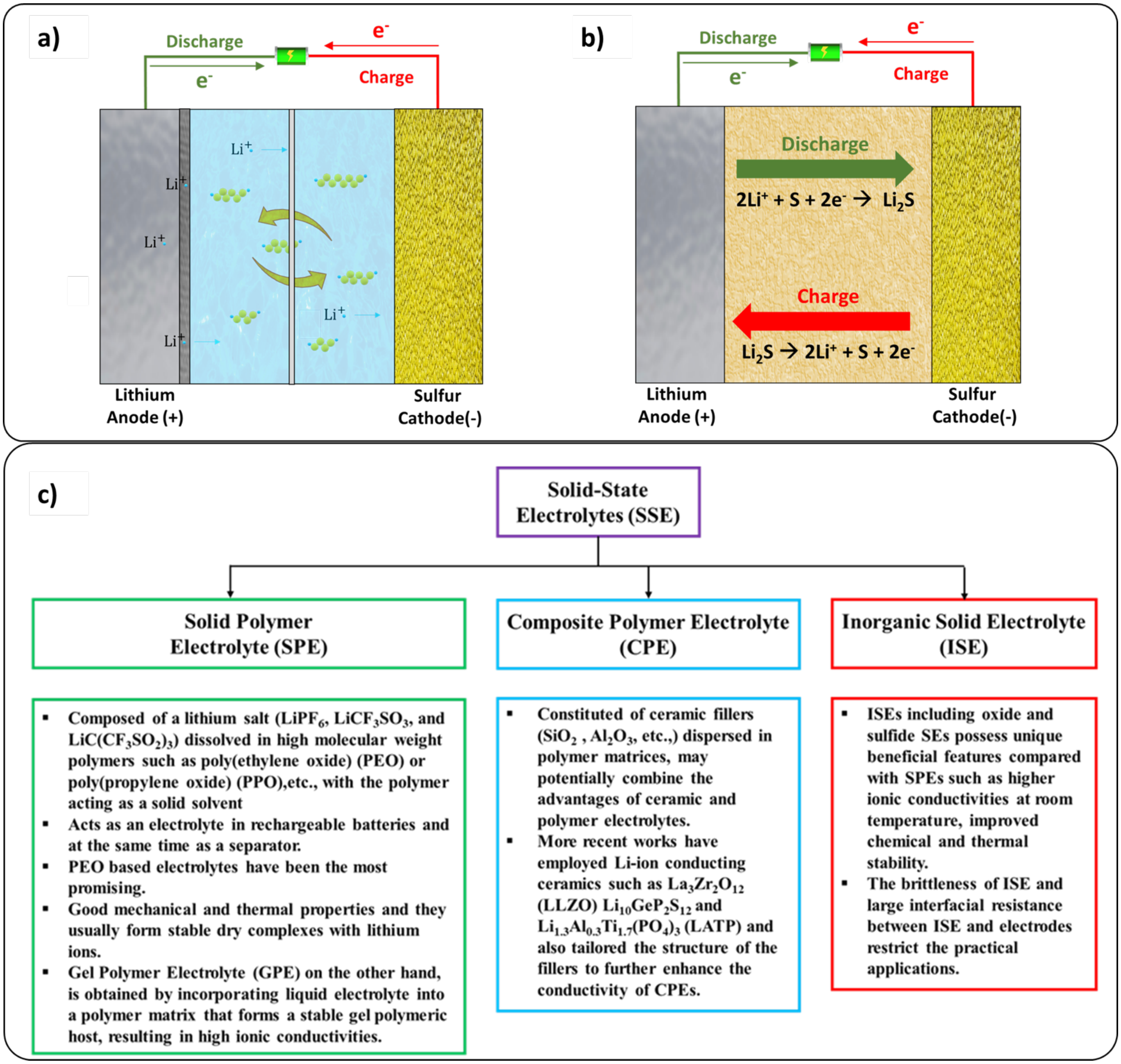

**Figure 1**. Illustrations of Lithium-Sulfur (Li-S) Batteries employing a) liquid and b) polymer-based solid electrolytes and c) Attributes of different solid-state electrolyte materials for Li-S batteries.

## 2. Polymer-based solid electrolytes for Li-S batteries challenges

Li-S batteries employing polymer-based electrolytes are viable contenders for future energy storage systems; however, the following challenges must be addressed for their practical implementation.

*2.1 Low Ionic Conductivity*

The ion conductivity of SSEs at room temperature remains low compared to liquid electrolytes, leading to suboptimal cell electrochemical performance. The electrolytes should simultaneously exhibit high ionic conductivity and electron transport insulation to prevent the formation of lithium dendrites and mitigate self-discharge. Ion transport in polymers is ascribed to segmental motions of the polymer chain, which facilitate cation transport within the polymer backbone and between polymer chains. Amorphous polymers offer higher degrees of freedom of the segmental motions. Polymers must also exhibit high solubility for anion salts, limiting the choice of polymers to high dielectric constant materials[40,41]. Inorganic filler particles are known to increment the amorphous fraction in polymers such as PEO, which exhibit simultaneously both amorphous and crystalline order. Filler particle properties are also chosen to increment Lewis-acid base interactions to promote anion-salt dissociation[42]. A reduction of polymer crystallinity can be achieved through polymer blending, copolymerization, and filler addition[43,44]. Whereas ISEs exhibit higher ionic conductivities than other SSEs, they are brittle and challenging to incorporate in battery devices to produce low-ionic resistance interfaces with robust mechanical properties; furthermore, they are costly to synthesize. On the other hand, the combination of ISE nanoparticles and polymer electrolytes is employed in composite polymer electrolytes to significantly increment the ionic conductivity of the polymer matrix[45]. In addition, incrementing the cation transference number from current values of ~0.2 – 0.4 to 1.0 would significantly improve current battery performance characteristics[46]. Incorporation of ceramic active fillers such as Lithium lanthanum zirconium oxide (LLZO)[46], Lithium aluminium titanium phosphate (LATP), etc., into the polymer-based solid electrolytes, has been reported to increase the lithium transference number[47]. Polymer-based solid electrolytes provide transport for both cations and the anion salts. The mobility of $Li^+$ is higher than the anions on account of size & solvation effects and ionic interactions[48]. Significant polarisation and compositional gradients develop over the process of charge and discharge cycling, particularly at high rates, hindering battery performance and charge/discharge rates. To circumvent said gradients, active research on anion Lithium salts, polymer additives[49,50], and ionic liquids[51,52] is an active area of investigation.

*2.2 Polysulfide shuttle effect*

The shuttle effect in liquid electrolyte-based Li-S batteries is illustrated in Figure 2a. Typical discharge and charge profiles are presented in Figure 2b. Two distinct discharge plateaus can be observed; within the upper discharge plateau, the $S_8$ undergoes reduction to $S_8^{2-}$, followed by $S_6^{2-}$ and $S_4^{2-}$, with an average voltage of 2.3 V (compared to $Li/Li^+$). This corresponds to a theoretical capacity of 418 $mAh.g^{-1}$. The discharge plateau at 2.1 V (vs $Li/Li^+$) represents the continued reduction of polysulfides to insoluble $Li_2S_2/Li_2S$, resulting in a theoretical capacity of 1254 $mAh.g^{-1}$. In the process of subsequent charging, the conversion of $Li_2S$ to elemental sulfur proceeds through the development of intermediate polysulfides, leading to a reversible cycle[53–56].

The discharge characteristics of Li-S batteries with polymer-based electrolytes can differ from those with liquid electrolytes. Polymer-based electrolytes can be designed to hinder the shuttle effect caused by polysulfide diffusion. This can reduce and potentially eliminate the amount of polysulfides reaching the active electrodes, resulting in enhanced stability and battery capacity retention. Evidence of polymer electrolytes exhibiting a single discharging plateau is reported in refs. [57–59]. The solid-state nature of electrolytes based on PEO limits the diffusion and transport between the active electrodes. Polysulfides can be effectively immobilized within the polymer matrix through strong chemical and electrostatic interactions with PEO. In addition, the solubility of Polysulfides in the PEO matrix is significantly lower than in liquid electrolytes, further restricting their mobility[60]. Porous polymer morphologies, as in the case of PVDF-based electrolytes, enable the trapping of dissolved polysulfides, and the polymer network can partially limit the movement of polysulfides[61]. Polymer electrolytes aid in reducing the shuttle effect in Li-S batteries by employing different mechanisms, including creating a stable solid electrolyte interface, physically encapsulating the sulfur or lithium polysulfides, and preventing contact between the lithium polysulfides and the anode. The choice of polymer chemical composition and structure can be engineered to minimize and, in principle, eliminate the shuttle effect.

*2.3 Electrode interfaces*

The interface properties (adhesion, chemical, mechanical robustness, low ion resistance) between the lithium metal anode and SSEs are critical for battery performance. Whereas liquid electrolytes readily wet the electrode interface, this is not true for SSEs. The large cathode volume changes during lithiation and delithiation can compromise the mechanical integrity of the interfaces. Furthermore, SSEs in contact with Li cations can undergo chemical/electrochemical reactions leading to their decomposition and the formation of by-products at the interface[62]. The solid polymer electrolyte/anode interface is less resilient than the case of flexible GPE/anode materials. Potential solutions to enhance interface contact in solid polymer electrolytes include incorporating intermediate buffer layers [63] and applying coating materials to provide surface passivation protection and inhibit the diffusion of undesirable reaction products. When the thickness of the coating material is less than that of the decomposition product layer, it can effectively reduce the interface resistance[64]. The growth of lithium dendrites, which primarily arises from non-uniform Li deposition at the interface, is also a limiting factor[65]. The high mechanical strength of polymer-based solid electrolytes hinders the growth of dendrites. However, if the SSE exhibits electronic conductivity, it facilitates the binding of $Li^+$ ions to electrons within the electrolyte, gradually forming lithium dendrites. Thus, several strategies for inhibiting lithium dendrites have been proposed, and a deeper understanding of the mechanisms underpinning dendrite formation in SSEs is needed to eliminate their formation[66].

*2.4 Mechanical robustness and safety*

The sulfur cathode experiences substantial changes in volume during the cycling process, with expansion reaching up to 80%[67]. To ensure adequate contact with the electrodes and prevent the occurrence of delamination or cracking, polymer electrolytes need to exhibit high mechanical strength and be flexible. The high mechanical strength can effectively inhibit the formation of lithium dendrites, as a high shear modulus offers superior mechanical resistance to dendrite penetration. Solid polymer electrolytes comprise non-combustible constituents; thereby, they are intrinsically safer than liquid electrolytes. A mechanically robust polymer electrolyte also serves as a physical barrier to the movement of polysulfides, thereby reducing the shuttle effect[68,69].

*2.5 SSEs materials processing*

A scalable, rapid, low-cost manufacturing process for polymer-based electrolytes is required; in addition, synthesis methods define the structure-property relationships in SSEs[70]. In the case of composite polymer electrolytes, the polymer, anion salts and filler particles need to be homogeneously mixed, dried, and cast to form thin film membranes. Fabrication of pinhole-free SSEs thin films can be realized by casting, extrusion, or hot-pressing methods. In the case of composite polymer electrolytes, particle size reduction of the inorganic filler particles significantly affects the CPE ionic conductivity. It can be achieved by incorporating ball milling into the CPE fabrication process[71].

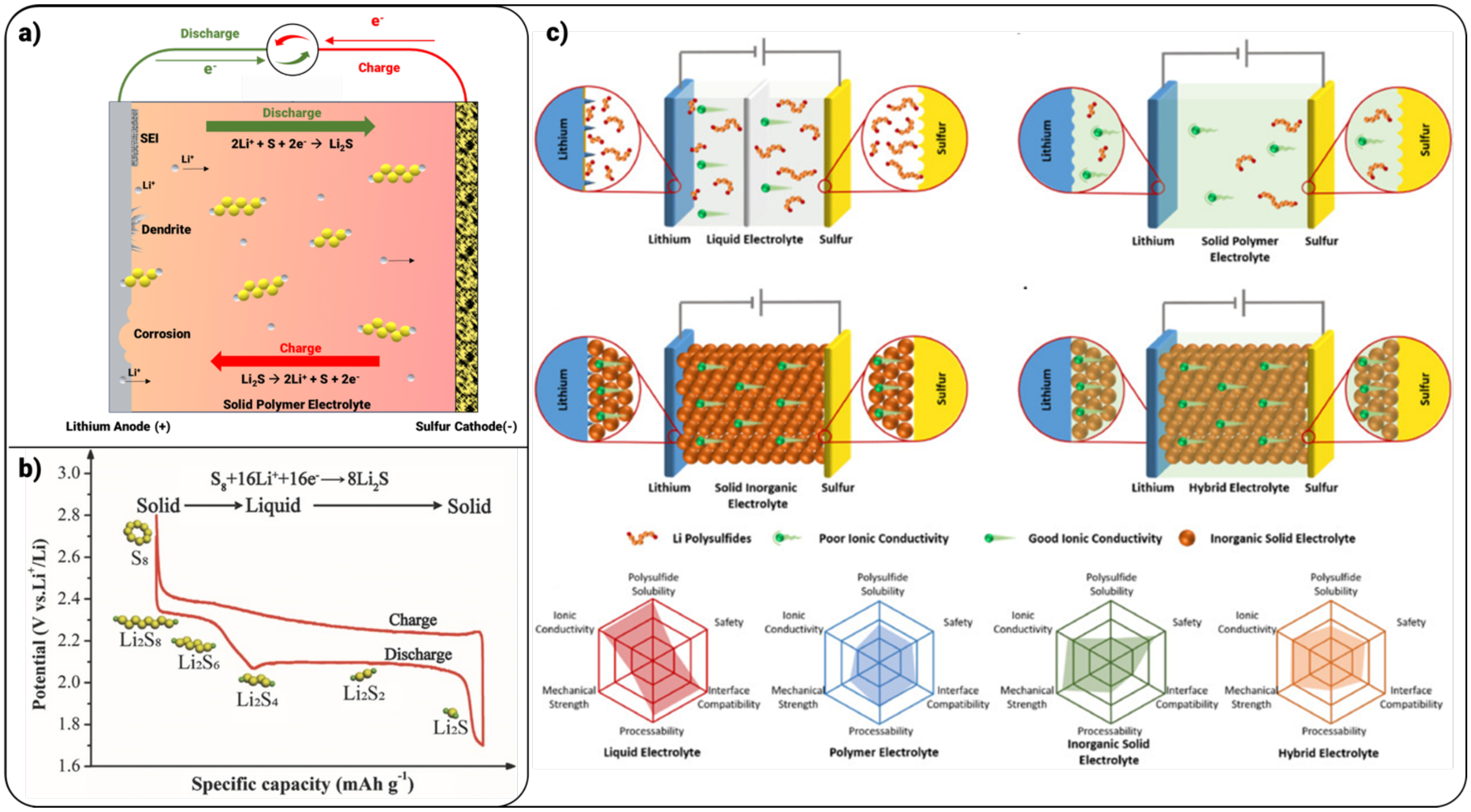

**Figure 2**. a) Schematic representation of the shuttle effect of polysulfides and dendrite formation during the operation of Li-S batteries employing liquid electrolytes. b) discharge and charge profiles of a typical Li-S Battery. Reproduced with permission.[55] Copyright 2017, John Wiley and Sons. c) Illustration of different electrolytes in Li-S Batteries and their radar maps of battery performance attributes. Reproduced with permission. [72] Copyright 2019, Elsevier.

**Table 1.** Advantages, disadvantages, and potential solutions for solid-state electrolytes.

| Electrolyte Type | Advantages | Disadvantages | Strategies |
| --- | --- | --- | --- |
| Solid Polymer Electrolyte (SPE) | Good flexibility.<br>Excellent Compatibility.<br>Easy to process.<br>Low Cost. | Low ionic conductivity.<br>Minimal mechanical strength.<br>Poor thermal stability.<br>Narrow electrochemical window. | Blending.<br>Copolymerization.<br>Cross-linking.<br>Establishing fast ion pathways. |
| Composite Polymer Electrolyte (CPE) | High ionic conductivity.<br>Suitable mechanical properties.<br>Low interfacial impedance. | Inorganic filler aggregation.<br>Intricate operation process. | Control packing concentration and particle size.<br>Modification of fillers. |
| Inorganic Solid Electrolyte (ISE) | High ionic conductivity.<br>High voltage resistant.<br>High safety standards.<br>Brilliant chemical and electrochemical stability.<br>Mechanical properties. | Poor interface compatibility.<br>Brittleness and formation of cracks.<br>Poor air stability.<br>Poor flexibility.<br>High cost. | Interface modification.<br>Putting intermediate layer.<br>Metal oxide doping. |

**3. Preparation methods of polymer-based electrolytes.**

Composite polymer electrolytes comprise inorganic active fillers and anion salts embedded in the polymer matrix. For Li-S batteries, the inorganic active fillers can be designed to anchor polysulfides, suppressing the shuttle effect. The interface between the composite SSE and the electrodes must be engineered to minimize interface resistance and provide mechanical strength. Furthermore, inorganic fillers contribute to lithium-ion transport in the polymer matrix; thus, integrating several functional materials into SSE provides a route to engineer ion transport and mechanical properties of SSEs[58]. As mentioned, the polymer matrix must efficiently dissolve the anion lithium salts while retaining their negative charge and exhibiting chemical and thermal stability[73].

The most popular SPE fabrication technique is solution casting, which involves dissolving the polymer and its constituents in a suitable solvent. Mixing and filler particle size reduction can be achieved by ball milling. The mixture is poured onto a mould and annealed to allow evaporation of volatile solvents; after cutting and slicing the CPE, it is integrated with the electrodes to form a battery[74,75]. The solvent must be nonaqueous and dissolve both the polymer and lithium salts. In addition to PEO, SPEs based on other polymers such as PVDF[76], poly(vinylpyrrolidone)(PVP)[75], PMMA[77], polyvinyl alcohol (PVA)[78], etc., have been employed.

Different preparation methods for CPEs are illustrated in Figure 3. Various fillers such as Lithium lanthanum zirconium oxide (LLZO)[46,79], Lithium aluminium titanium phosphate (LATP)[80], $Li_{10}SnP_2S_{12}$[81], $MnO_2$[82], $TiO_2$[83], halloysite nanotubes[84], sodium superionic conductors (NASICON)[57], and Ca-$CeO_2$[85], $Al_2O_3$[86] etc., have been employed in the preparation of CPEs. Li *et al.*[87] incorporated $Li_{10}SnP_2S_{12}$ into PEO/LiTFSI matrixes using the solution casting method. In another example, following 24 hours of ball milling, a uniform mixture of PEO, $LiCF_3SO_3$, $Li_2S$, and $ZrO_2$ was hot pressed to produce a firm film. The PEO- $LiCF_3SO_3$ complex in this electrolyte is the $Li^+$ transport medium. $ZrO_2$ fillers served as stabilizers at the lithium anode and electrolyte interface, while the $Li_2S$ filler offered sulfide dissolution prevention[88]. The drop coating method has been used to fabricate sandwich-type composite electrolytes to enhance interfacial compatibility and $Li^+$ ionic conductivity. Wang *et al.*[89] developed a sandwiched NASICON-type LATP solid-state electrolyte with surface modifications provided by an SPE. To fabricate the SPE membrane onto the LATP surface, an SPE precursor solution was applied onto one side of the LATP ceramic pellet using the drop coating technique. The application of in situ ultraviolet (UV) curing was used to prepare CPEs, facilitating the uniform dispersion of fillers within the polymer matrix. UV curing requires unsaturated polymer monomers and photoinitiators[90]. The photoinitiator commonly employed is 2-hydroxy-2-methyl-1-phenyl-1-propanone (HMPP)[91]. The typical monomers utilized include poly(ethylene glycol) diacrylate (PEGDA), butyl acrylate (BA), ethoxylated trimethylolpropane triacrylate (ETPTA) trimethylolpropane propoxylate triacrylate (TPPTA), and di(trimethylolpropane) tetraacrylate (DTPTA), among others[92]. In addition, electrospinning[93], phase inversion[94], spin coating[72], melt intercalation[95], and dip coating[96] techniques are also employed for the fabrication of CPEs[97].

The solution casting method provides consistency in thickness control of membranes. It has the potential to provide a homogeneous dispersion of anion salts and filler particles in the polymer matrix. It requires the optimal selection of solvents for polymer, salts, and filler dissolution[98]. Hot pressing offers a method to achieve strong bonding between electrode interfaces. The use of hot pressing can improve ion mobility by decreasing the crystallinity of the polymer. On the other hand, drop coating processing may result in variations in film thickness and anion salt and filler particle distribution in the SSE. This can compromise the electrolyte ionic conductivity. Effective curing improves the connectivity of the polymer network, therefore improving ion transport. Curing techniques can potentially enhance the mechanical properties of the SSE via the formation of stiff networks[33,35,99,100].

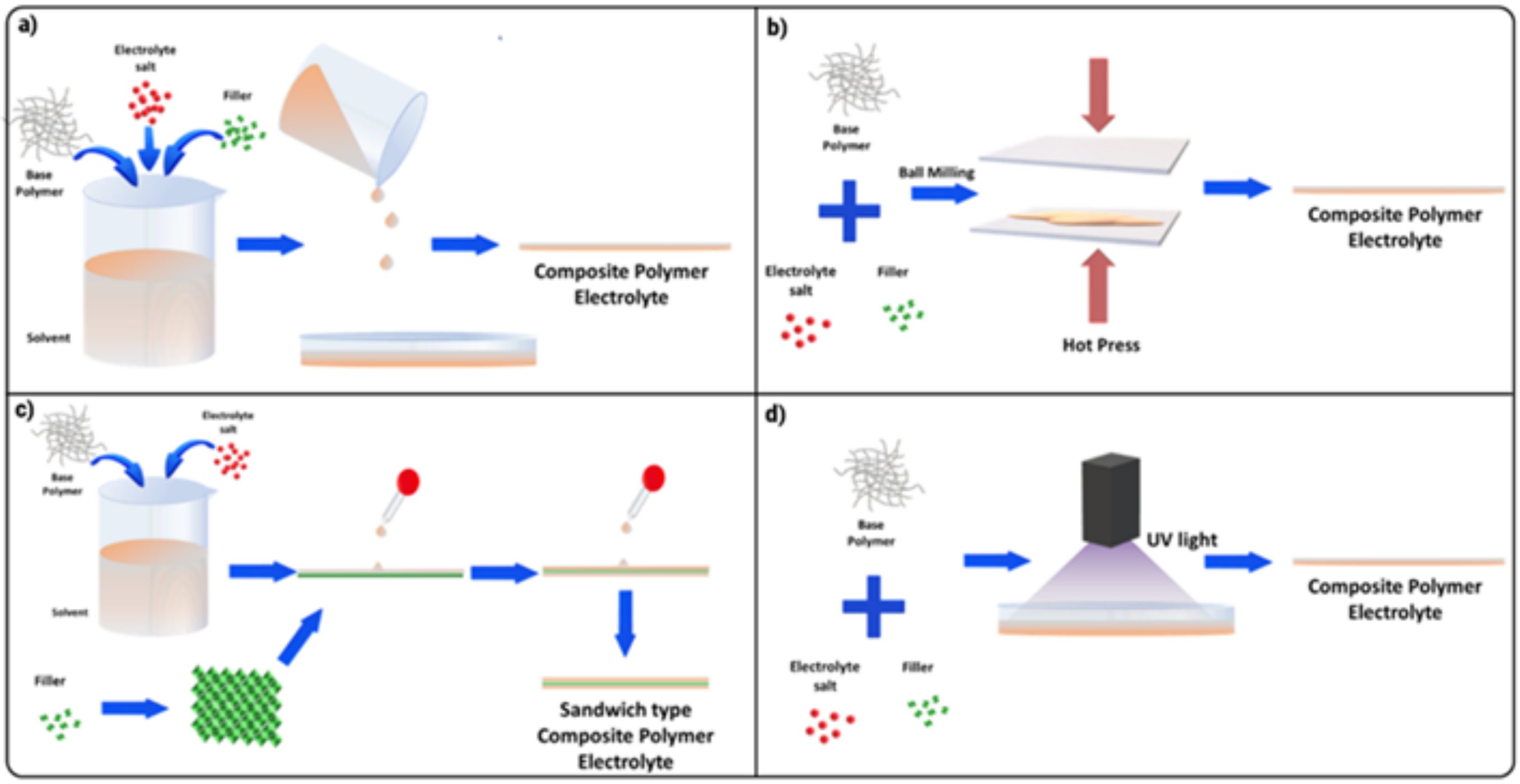

**Figure 3**. Schematics of various preparation techniques for fabricating composite polymer electrolytes: a) Solution Casting, b) Hot Press, c) Drop coating, and d) UV curing.

**4. Recent Advances in Polymer-based electrolytes for Li-S batteries.**

Polymer-based solid electrolytes offer advantages over inorganic counterparts and are amenable to facile, low-cost fabrication. Utilising polymer electrolytes can effectively prevent the depletion of active materials caused by the diffusion and secondary reactions of polysulfides in liquid electrolytes, thereby mitigating the shuttle effect. The growth of lithium dendrites can be suppressed in polymer electrolytes by modification of the surface energy at the lithium metal anode/SSE interface to prevent heterogeneous Li-metal nucleation and by providing a high shear modulus barrier for dendrite growth. Additional properties that must be met by SPEs to meet Li-S battery requirements include providing high ionic conductivity and cation transference number. In SPEs, the polymer and the lithium salts combine to form complexes and ion transport is driven by the segmental motions of the polymer chains, which facilitate the movement of anions and cations to nearby coordination sites in the polymer chains. The amorphous regions of the semicrystalline polymers are mainly responsible for ion transport; thus, ion-conducting polymers should exhibit amorphous or semicrystalline structures at room temperature and have high dielectric constants to dissolve the anion salts [101,102]. Polymers that contain a high concentration of polar functional groups, such as –O– and –C═O, exhibit excellent solubility towards lithium salts, a crucial requirement for an electrolyte material[103]. The polymer must exhibit chemical stability with electrode materials, and in the presence of polysulfides, it must also exhibit electrochemical stability across a broad range of voltages.

Whereas several SPE materials readily meet stability and safety criteria, their suboptimal ionic conductivity and some of their mechanical characteristics require improvements. This section reviews recent results and advances in polymer-based solid electrolyte systems to address these bottlenecks.

*4.1 PEO-based electrolytes*

PEO-based materials are promising polymer hosts for solid-state electrolytes. Their lithium-ion conductivity, resulting from coordination interactions between lithium cations and the ether oxygen groups, together with their semicrystalline nature, provides high ionic conductivity in comparison to other polymers. However, the ionic conductivity is inadequate for battery applications and one of the reasons is their high degree of crystallinity, which restricts segmental motions. Thus, reducing the crystalline fraction has been the subject of several recent studies. Wei *et al.*[104] incorporated fullerenes ($C_{60}$) to produce a CPE with high ionic conductivity, electrochemical stability, and safety compatibility. With the addition of $C_{60}$, the ionic conductivity increased from $5.7 \times 10^{-5}$ to $1.27 \times 10^{-4}$ S cm$^{-1}$ at 60 °C, which was attributed to the reduction in crystallinity. The elevated interface energy of $C_{60}$ towards Li obstructs the proliferation and infiltration of Li dendrites and contributes to cycling stability for 1000 hours. The advantages of Li-S batteries with SPE made of PEO and 1% $C_{60}$ are illustrated in Figure 4a. Increasing the liquid-solid conversion rate (LPSs → $Li_2S_2$/$Li_2S$) at the cathode-electrolyte interface, the $C_{60}$ additive decreased the shuttle effect. The $C_{60}$ addition increased the polymer segmental motions and dissociated the ion association of LiTFSI. In addition, the higher interface energy between $C_{60}$ and lithium metal can effectively block the growth and penetration of dendrites. Liu *et al.*[105] integrated a bifunctional additive $Nb_2CT_x$ MXene into PEO/LiTFSI-based polymer electrolytes. A $Li^+$ transference number of 0.37 at 60 °C was obtained by adjusting the $Nb_2CT_x$ MXene sheet size from 500-300 nm to below 100 nm. The ionic conductivity of the membrane was 2.62 x $10^{-4}$ Scm$^{-1}$, whereas the ionic conductivity for the PEO/LiTFSI membrane without the MXene was 8.6 x $10^{-5}$ Scm$^{-1}$. The incorporation of $Nb_2CT_x$ MXene resulted in enhanced $Li^+$ conductivity and polysulfide absorptivity. Figure 4b schematically shows the effect of the MXene in stabilizing the polysulfides. Li *et al.*[106] designed a CPE consisting of polyethylene glycol azobenzamide (PGA) and PEO with added LiTFSI, N-methyl-n-propylpyrrolidinium bis(trifluoromethanesulfonyl)imide ($Py_{13}$TFSI) and $Al_2O_3$ fillers using the solution casting method. PGA facilitates the dissociation of lithium salts, leading to an increase in free lithium ions, thus enhancing the ionic conductivity. A significant decrease in the shuttle effect was also reported and attributed to the high affinity between PGA and the polysulfides (Figure 4c). Li solid-state NMR spectroscopy and density functional theory (DFT) were utilized in the study. The ionic conductivity in these CPEs was measured to be $4.9 \times 10^{-4}$ S cm$^{-1}$, and the $Li^+$ transference number was 0.53 at 50 °C. Grotkopp *et al.*[107] employed a solvent-free tape casting method to fabricate PEO-based CPEs containing silicon dioxide ($SiO_2$) fillers. The CPE films were cast onto Li metal anodes, and it was found that the PEO-based films protected the Li-metal surface, resulting in improved stability and consistency. A current density of 400 mAh.gs$^{-1}$ after 250 cycles was measured.

In contrast, the bare Li-metal electrode achieved a current density of 200 mAh.g$^{-1}$ after 250 cycles. To enhance the mechanical strength and thermal stability of the PEO polymer matrix and to hinder crystallization, Li *et al.*[108], incorporated a polytetrafluoroethylene (PTFE) skeleton and $Li_7La_3Zr_2O_{12}$ (LLZO) fillers. The multi-functional effects of the prepared electrolyte in a lithium-sulfur battery are shown in Figure 4d. The PTFE@LLZO@PEO electrolyte demonstrated suppression of the shuttle effect and reduction of interfacial impedance between the electrolyte and the cathode while also promoting strong adhesion between the electrolyte and the electrode. Furthermore, the PTFE@LLZO@PEO electrolyte led to a homogeneous distribution of $Li^+$ at the anode interface, thus impeding the formation of lithium dendrites. The PTFE membrane exhibited a maximum tensile strength and strain of 3.7 MPa and 35.33%, respectively. Upon loading with LLZO@PEO, the tensile strength and strain incremented to 4.82 MPa and 73.33%, respectively. This CPE exhibited thermal stability up to 375 °C. A novel PEO-PAN-LLZO composite gel polymer electrolyte (GPE) was designed by Xie *et al.*[109], the composite GPE was added to a commercial liquid electrolyte (1.0 wt% Lithium nitrate ($LiNO_3$), 1 M LiTFSI Dimethoxyethane (DME)/Dioxolane (DOL) (1:1)) to integrate the benefits of liquid-state and solid-state electrolytes. This GPE was intended to exhibit superior interface wettability, high ionic conductivity, and excellent mechanical properties while maintaining a high level of safety. The membrane was coated with a conductive acetylene black layer to impede the migration of polysulfides and improve their redox kinetics. The PEO-PAN-LLZO GPE exhibited an increased maximum tensile strength of 10.64 MPa, whereas the PEO-PAN GPE without the LLZO additive showed a tensile strength value of only 5.40 MPa. These improvements were attributed to the LLZO, which effectively absorbs a significant portion of the deformation energy. Zhang *et al.*[110] incorporated nano $In_2O_3$ fillers into PEO/LiTFSI polymer electrolytes, enhancing ionic conductivity. It also created a Li-In alloy layer at the polymer/anode interface, accelerating the diffusion of lithium ions and hindering side reactions between the lithium metal anode and the electrolyte. At 60 °C, the sample containing 12% wt. load of $In_2O_3$ exhibited an ionic conductivity of 5.27 x $10^{-4}$ S cm$^{-1}$. Sheng *et al.*[34]constructed a membrane electrolyte composed of a copolymer of PEO and PAN, wherein the PAN fibres served the dual purpose of a filler and cross-linking agent. This architecture offered not only enhanced ionic conductivity, elevated mechanical robustness, and effective prevention of lithium dendrite formation but also impeded polysulfide shuttling due to the potent adsorption of polysulfide by the C=N–O functional groups generated during the cross-linking process of PEO and PAN. Inactive fillers have also been shown to diminish the shuttle effect in Li-S batteries based on PEO. Zhuoran *et al.*[111] synthesized and incorporated NiO/$C_3N_4$ heterojunction structures into PEO. The combination of NiO and $C_3N_4$ was selected because of the ability of $C_3N_4$ to provide pathways for the movement of lithium ions and the effective adsorption of LiPSs by NiO. Built-in electric fields are present at the interface between NiO and $C_3N_4$. The $Li^+$ cation and $TFSI^-$ anion in lithium salts can rapidly dissociate when subjected to an electric field, generating additional free lithium ions. The NiO/$C_3N_4$ heterojunction doped PEO-based polymer electrolyte exhibited an ionic conductivity of $1.2 \times 10^{-4}$ S

cm$^{-1}$ with a $Li^+$ transference number of 0.25 at 30 °C. To reduce PEO crystallinity, increase its ionic conductivity and enhance its mechanical strength, Ji *et al.*[112] developed a PEO solid electrolyte with a cellulose acetate nanofiber network modified by anion-exchanged cations. The introduction of quaternary ammonium cation-modified cellulose disrupted the arrangement of the PEO molecular chains, thereby reducing crystallinity and increasing ionic conductivity to $2.07 \times 10^{-4}$ S cm$^{-1}$ at 60 °C. The quaternary ammonium groups in the cellulose molecular structure anchored the TFSI-anions via Lewis acid-base interactions, enhancing lithium-ion migration. These quaternary ammonium groups are also bound to polysulfides, preventing their mobility. The inclusion of a three-dimensional nanofiber mesh obtained from TFSI- quaternary ammonium cation-modified cellulose enhanced the mechanical strength of the PEO electrolyte (from 0.49 to 7.50 MPa), effectively addressing safety issues related to the infiltration of lithium dendrites.

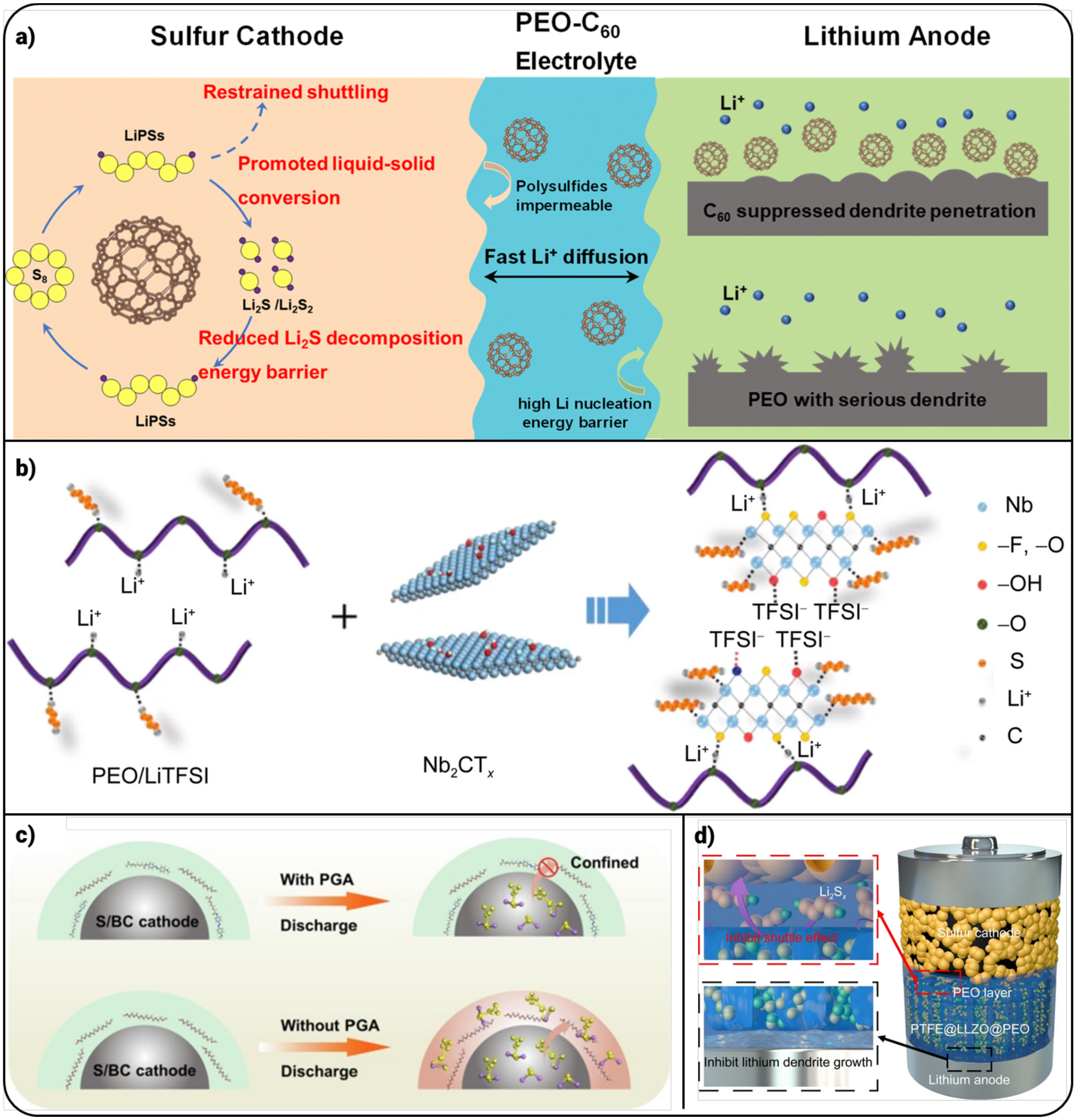

**Figure 4.** a) Structure and advantages of Li-S batteries with SPE made of PEO and 1% $C_{60}$. Reproduced with permission.[104]. Copyright 2023, Royal Society of Chemistry. b) Illustration of improving lithium transport and suppressing polysulfide shuttling in $Nb_2CT_x$ MXene PEO-based polymer electrolyte. Reproduced with permission.[105] Copyright 2023, Springer Nature. c) Proposed mechanism of S/BC cathodes in electrolytes with or without PGA. Reproduced with permission.[106] Copyright 2023, Elsevier. d) Illustration of the beneficial attributes of PTFE/LLZO/PEO electrolyte in a Li-S battery. Reproduced with permission. [108] Copyright 2022, Springer Nature.

*4.2 PVDF-based electrolytes*

PVDF and its derivatives are extensively employed as polymer hosts because of their exceptional mechanical and electrochemical characteristics. The ease of film preparation also renders it a prime option for polymer electrolyte applications. PVDF exhibits a high dielectric constant of 7–13 at 1 kHz[113], whereas it is around 5 for PEO. Additionally, PVDF offers ease of processing and exceptional thermal and electrochemical stability. Castillo *et al.*[114] developed a PVDF-based GPE, a copolymer of PVDF with hexafluoropropylene (PVDF-HFP). To address inadequate ionic conductivity at lower temperatures, this GPE was plasticized with poly(ethylene glycol dimethyl ether) (PEGDME) and tetraethylene glycol dimethyl ether (TEGDME) to achieve an ionic conductivity value of around $2.7 \times 10^{-4}$ S $cm^{-1}$ at room temperature. Leng *et al.*[115], through a single-step electrospinning technique and heat treatment, synthesized a bilayer PAN/CB/VOOH-PAN/PVDF(HFP) separator. Each layer provided distinct benefits: the PAN/CB/VOOH layer, located proximal to the cathode of the Li-S battery, functioned as a conductive barrier layer, and the PAN/PVDF(HFP) layer provided high thermal stability. The polar VOOH adsorbed and catalysed lithium polysulfide species, thereby impeding the shuttle effect. Comparative studies were also conducted regarding thermal properties, $Li_2S_6$ diffusion and electrochemical performance between separators made from Celgard, PAN, PAN/CB/VOOH-PAN/PVDF(HFP) (PCVPP), and heat treated-PAN/CB/VOOH-PAN/PVDF(HFP) (HPCVPP). The results are presented in Figure 5. The findings suggest that the conductive blocking layer composed of PAN/CB/VOOH in the HPCVPP separator serves as a substrate for the LPS electrochemical reactions. Additionally, this polar layer absorbs and catalyses LPS species, impeding the shuttle effect. The electrolytic wettability of the HPCVPP separator was efficient enough to enhance the electrochemical performance of the Li-S battery. The cyclic voltammetric curves (Figure 5a-d) analysis reveals that the Li-S battery incorporating PCVPP and HPCVPP separators exhibited more pron ounced redox peaks and larger currents in the CV curve. This indicates that the inclusion of these new separators enhanced the electrochemical reaction in the battery due to the catalytic and adsorption effects of VOOH on $Li_2S_6$. The Nyquist plots (Figure 5e,f) indicate that the Li-S batteries employing PCVPP and HPCVPP separators exhibited superior ionic conductivity. This may be ascribed to their higher porosity and enhanced electrolytic absorption inside the three-dimensional nanofibrous architecture. After 500 charge-discharge cycles, the discharge capacity of the PAN, PCVPP, and HPCVPP separator-incorporated Li-S batteries reduced from around 680, 750 and 810 mAh.$g^{-1}$ to 313, 581, and 644 mAh.$g^{-1}$, respectively. Additionally, the capacity retention rate was found to be 45.2%, 69.0%, and 79.4%, respectively. The Li-S battery, which contained an HPCVPP separator, exhibited a discharge capacity degradation rate of 0.2% per charge-discharge cycle (Figure 5k). This may be related to the deposition of a minor proportion of lithium sulfide on the membrane, leading to partial adsorption and subsequent loss of catalytic sites during an extended cycle test. The charge-discharge specific capacity of the Celgard, PAN, PCVPP, and HPCVPP separator-incorporated Li-S batteries at various battery voltages and charge-discharge rates is shown in Figure 5g-j. The Li-S battery's specific discharge capacity progressively declined as the battery discharge rate increased. The increased specific capacity of the HPCVPP separator-incorporated Li-S batteries is ascribed to the porous nature of the separator, facilitating the retention of the electrolytes. This is because, with more surface area associated with the pores, the electrolyte can make more contact, and the separator can retain more electrolytes in the pores. Additionally, the first layer (PAN/CB/VOOH) of the HPCVPP separator serves as a site for electrochemical reactions of LIPS species to initiate the activation of dead sulfur (lower-order PSs exhibit low solubility and tend to precipitate out of the electrolyte resulting in the formation of "dead" sulfide species on the surfaces of the electrodes) and prevent battery capacity degradation resulting from the deactivated materials within the battery. Zhu *et al.*[42] investigated the role of electrolyte/filler interfaces (EFI) in PVDF-HFP-based CPEs to enhance $Li^+$ ion transport by incorporating Prussian blue analogues (PBA) fillers, $Co_3[Co(CN)_6]_2 \cdot 12H_2O$ with and without Co–O ligand modification, namely unsaturated coordination PBA (UCPBA) and saturated coordination PBA (SCPBA), respectively. The study revealed that the interactions between fillers and the polymer matrix play a crucial role in determining the overall performance of CPEs. It was found that the dissociation of $Li^+$ from LiTFSI and the diffusion of $Li^+$ through the EFI were facilitated by the absorption of $TFSI^-$ anions on compatible interfaces between UCPBA and LiTFSI. The batteries demonstrated improved performance metrics, including room-temperature ionic conductivity up to $3.6 \times 10^{-4}$ S $cm^{-1}$ (approximately three times that of filler-free composite) lithium transference number of 0.6.

Metalorganic frameworks (MOFs) are a distinct category of materials that comprise small pore dimensions and small particle sizes, providing large surface areas. Porous materials are attractive host materials as they inhibit volume change processes during lithiation. These characteristics make them ideal for various applications, such as ionic conductors or fillers in different solid polymer electrolytes[116,117]. Hence to utilize the advantages of MOFs, Wang *et al.*[36] developed a polymer electrolyte based on PVDF-HFP with incorporated $NH_2$-MIL-53(Al)-MOF fillers to reduce the polymer crystallinity. A solid composite electrolyte containing 10 wt% MOF, 30 wt% LiTFSI, 54.5 wt% PVDF-HFP and 5.5 wt% PMMA yielded an ionic conductivity of $5.54 \times 10^{-4}$ S $cm^{-1}$. The presence of PMMA likely reduced the interfacial resistance between the electrodes and the composite electrolyte, resulting in enhanced $Li^+$ cation transfer.

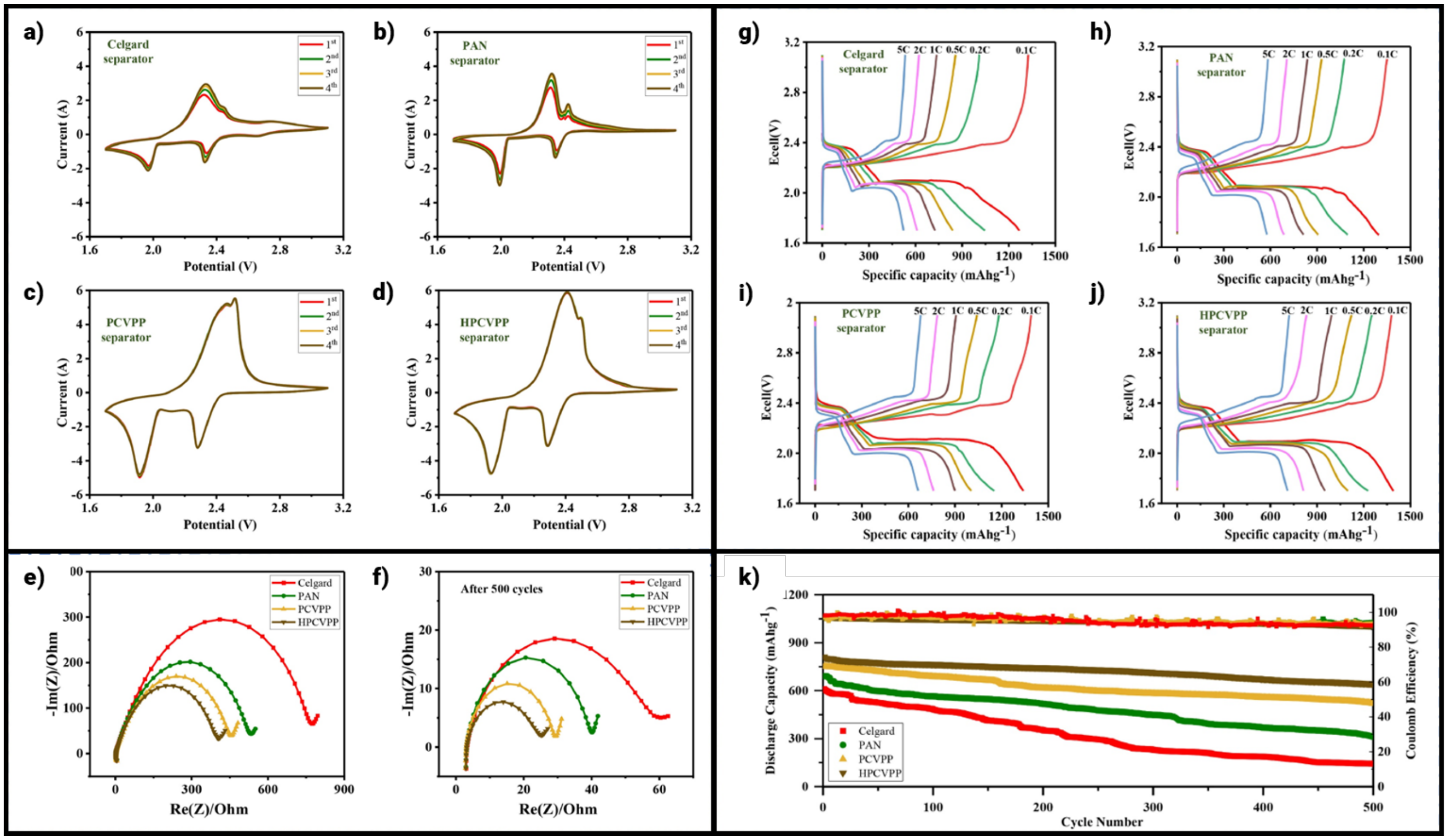

**Figure 5.** Voltammetry plots for Li-S batteries with different separators: a) Celgard, b) PAN, c) PCVPP, and d) HPCVPP; Nyquist plots of Li-S battery with different separators (Celgard, PAN, PCVPP, and HPCVPP), e) before and f) after 500 charge-discharge cycles. g) The Li-S battery's discharge performance (also including the Coulomb efficiency) evaluated for 500 charge-discharge cycles at a rate of 2C while using these separators. Charge–discharge specific capacity of h) Celgard, b) PAN, c) PCVPP, and d) HPCVPP separator-incorporated Li–S batteries under charge-discharge rates. Reproduced with permission.[115] Copyright 2023, Elsevier.

### *4.3. PMMA-based electrolytes*

PMMA based polymer electrolytes have attracted significant attention as they are easy to synthesize, have a low mass density, exhibit strong mechanical stability, low binding energy with ionic salts, and are generally more amorphous than PEO at room temperature[118]. A PMMA/Polypropylene carbonate (PPC)/ $LiNO_3$ composite was developed by Lu *et al*.[119] and found to act as an artificial solid electrolyte interphase that stabilizes the interface between the lithium metal and the electrolyte (1 M LiTFSI dissolved in DOL/DME). This composite solid electrolyte interphase (SEI) was mechanically stable and highly ionic conductive. This SEI provides C=O sites for anchoring polysulfides, which effectively limits the corrosion of the Li metal anode. In addition, when equipped with a 20 µm Li metal anode, Li-S batteries exhibited an initial capacity of 1166 mA h $g^{-1}$ at 0.5C. This SEI demonstrated enhanced structural uniformity and flexibility and could accommodate the volumetric changes of lithium metal. The outer layer of the SEI anchors the PSs, while lithium ions ($Li^+$) are swiftly transported through the SEI. Consequently, PMMA/PPC/$LiNO_3$ composite-SEI can efficiently lessen the utilization of Li metal, thus attaining elevated energy density.

Therefore, incorporating polycarbonate polymer/inorganic composite SEIs offers an approach to increase the energy density of Li-S batteries.

*4.4 Self-Healing polymer-based electrolytes.*

Self-healing ability, whereby a material can autonomously restore some or all its functionalities following mechanical impairments through the reconstruction of disrupted regions of the solid, is a desirable goal for battery electrode materials. Pei *et al*.[120] developed an integrated solid-state electrode/electrolyte utilizing a poly(ether-urethane) based electrolyte that displayed self-healing properties. At 30 °C, the ionic conductivity of polytetrahydrofuran (PTMG)- hexamethylene diisocyanate (HDI)- 2-hydroxyethyl disulfide BHDS/LiFSI (PTMG-HDI-BHDS/LiFSI) was measured to be $2.4 \times 10^{-4}$ S cm$^{-1}$, whereas that of PEO/LiFSI and PTMG-HDI/LiFSI were determined to be $1.2 \times 10^{-5}$ S cm$^{-1}$ and $6.5 \times 10^{-5}$ S cm$^{-1}$ respectively. Interfacial defects (that facilitate polysulfide shuttling, electrolyte eating, and unregulated dendritic development) at the electrolyte/electrode interface are reported to be *healed* in these structures, resulting in lower interfacial impedance, and the amorphous polymer framework contributes to the high ionic conductivity observed. It is suggested that the rearrangements of dynamic covalent disulfide and hydrogen bonds between the urethane groups in the poly(ether-urethane)-based SPEs contributed to the interfacial self-healing ability of the SPEs, allowing for the repair of solid/solid interfacial defects. Additionally, these groups facilitated the dissociation of lithium salts and enhanced the ionic conductivity[114]. Mong and Kim[121] reported the synthesis of a self-healable and flexible quasi-solid electrolyte employing poly(arylene ether sulfone) co-grafted with polyureidopyrimidinone and poly(ethylene glycol) (PAES-g-PU/2PEG) and PAES co-grafted with poly(1,2-bis(ureidoethylenemethacrylateethyl) disulfide) and PEG (PAES-g-PUS/2PEG). The self-healing capability of the compound is attributed to the presence of hydrogen and/or disulfide bonding (Figure 6c). The membranes were sliced in half to highlight the self-healing capability. Next, the dissected samples were carefully reattached, and the membranes exhibited structural rebuilding capabilities facilitated by the hydrogen and/or disulfide bonds. These electrolytes exhibited a remarkable self-healing efficiency of approximately 95.3% within a healing time of 5 minutes. The PAES-g-(PUS/2PEG) membranes that underwent a second self-healing process demonstrated a retention of 94.8% of their original properties. In addition, the flexible membranes exhibited a notable level of thermal stability up to 200 °C. In another work, Pei *et al*.[122] found that the rearrangements of dynamic covalent disulfide bonds and hydrogen bonds between urethane groups in the poly(ether-urethane)-based SPEs contributed to the exceptional interfacial self-healing ability of the SPEs, allowing for the repair of solid/solid interfacial defects throughout their entire lifespan. Additionally, these groups facilitated the dissociation of lithium salts and enhanced the ionic conductivity.

*4.5 Siloxane-based electrolytes*

Polysiloxane electrolytes exhibit exceptional thermal and electrochemical stability but poor ionic conductivities. To enhance their ionic conductivities, a high lithium salt concentration can be used or the implementation of a polymer matrix with a tailored molecular structure can be implemented through functional group modification. Introducing ionic groups to the siloxane backbone can potentially augment the transport of lithium ions inside the polymer matrix. Increasing the length of siloxane chains improves flexibility and elasticity, whereas reducing the length of the chains enhances ionic conductivity. The chain length in siloxane polymers can be engineered to meet these contradictory requirements[123]. Chen *et al*.[124] fabricated a polymer-in-salt polysiloxane SPE with bi-grafted polysiloxane (BPSO) copolymer, LiTFSI and PVDF. To achieve both desirable ionic conductivity and robust mechanical properties, the CPE incorporated a cellulose acetate matrix as a rigid substrate (Figure 6b). A tensile strength of 6.8 MPA, a cation transference number of 0.52 and an ionic conductivity of $7.8 \times 10^{-4}$ S $cm^{-1}$ at 25 °C was achieved. An innovative copolymer composed of PEO and a semi-interpenetrating polymer network has been developed using an in-situ cross-linking technique with siloxane-type ((bis(4-carbonyl benzene sulfonyl)imide -grafted siloxane) single-ion conductors (SICs) in PEO based polymer electrolyte for Li metal batteries by Hu *et al*.[125]. The resultant polymer electrolyte demonstrated ionic conductivity values of $1.10 \times 10^{-5}$ S $cm^{-1}$ at 25 °C and $1.41 \times 10^{-4}$ S $cm^{-1}$ at 60 °C. The flame retardancy of the electrolyte mitigates safety concerns arising from the combustion of electrolytes when Li-S batteries are operated at elevated temperatures. To develop an intrinsic flame-retardant siloxane ether-based electrolyte, Ma *et al*.[126] synthesized a LiFSI/tetrapropoxysilane electrolyte with sulfurized pyrolyzed poly(acrylonitrile) (SPAN) cathode. The characteristic high bond energy and thermal stability of the Siloxanes' Si-O bonds are responsible for the increased electrolyte flame resistance. The ionic conductivity of the electrolyte increased as the temperature increased and peaked at $3.75 \times 10^{-4}$ S $cm^{-1}$ at 80 °C.

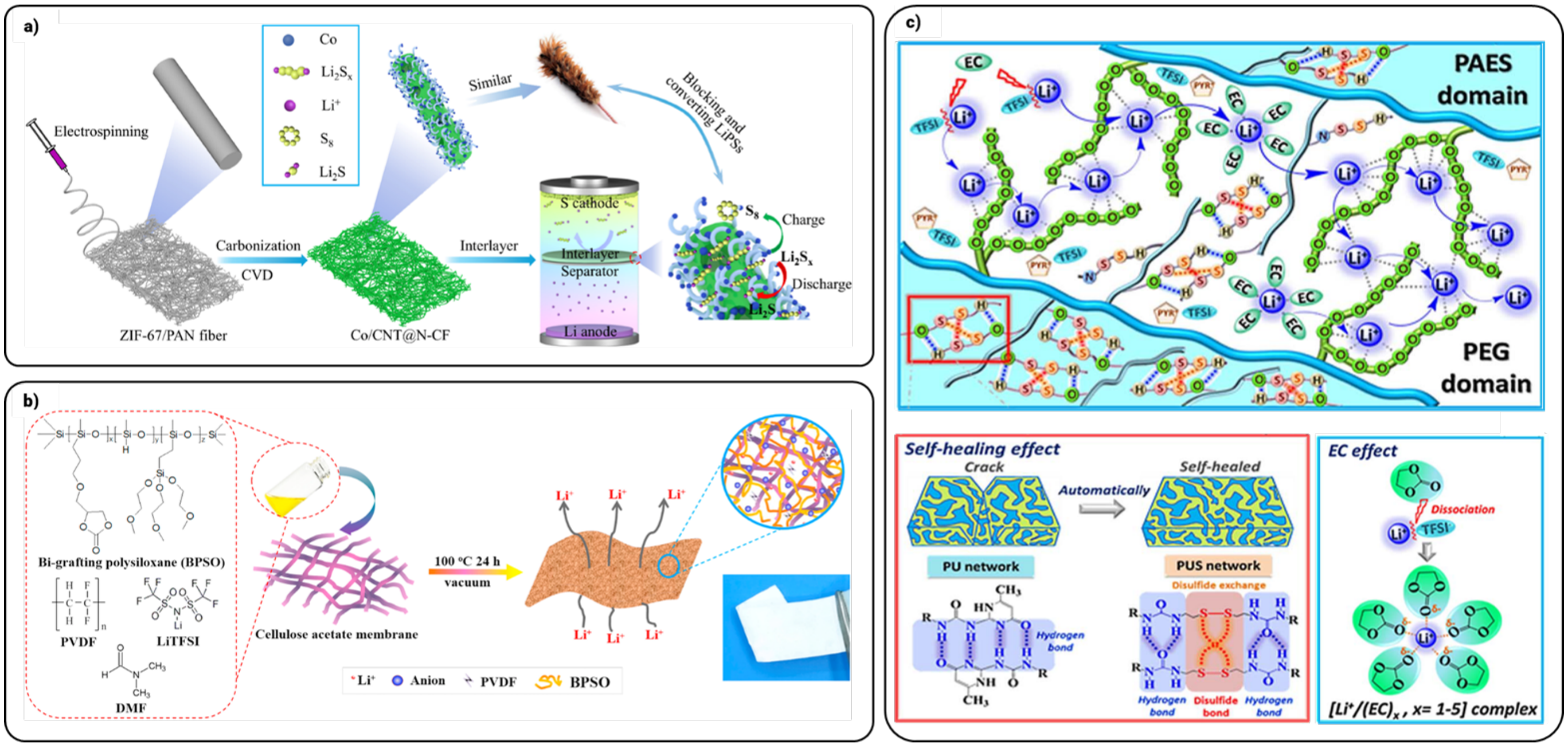

**Figure 6**. a) The fabrication process and operational mechanism of feather-like Co/CNT@N–CF interlayer. Reproduced with permission.[127] Copyright 2023,

Elsevier. b) Fabrication schematic of BPSO/PVDF/LiTFSI/ Cellulose acetate CPE membrane by solution-casting technique. Reproduced with permission.[124] Copyright 2018, Elsevier. c) Schematics of the formation of phase separation and the self-healing process of PAES-g-(PU/2PEG) and PAES-g-(PUS/2PEG) membranes through disulfide exchange reactions and hydrogen bonds. Reproduced with permission.[121] Copyright 2023, Royal Society of Chemistry.

## 5. Li-S solid-state polymer electrolyte efficiency improvements

SPEs for Li-S batteries should be lightweight and flexible, exhibit high mechanical strength, allow facile and low-cost processability, and provide chemical and electrochemical compatibility with the anode and cathode electrodes. SPEs must surpass ionic conductivities >$10^{-4}$ S $cm^{-1}$. PEO is the most widely utilized polymer in SPEs. However, its high degree of crystallinity and poor thermal and mechanical properties are impediments. Copolymerization, cross-linking, and blending techniques have been employed to engineer polymer properties and both organic and inorganic filler particles are employed to improve ion transport and polymer physical properties. This section discusses ongoing research to improve SPEs for Li-S battery applications.

*5.1 Ionic salts*

In SPEs for the Li-S batteries, the lithium salts and additives largely contribute to battery performance. LiTFSI is the predominant lithium salt used in Li batteries[128–130]. Other Lithium salts employed LiFSI, LiFTFSI, lithium (fluorosulfonyl) (pentafluoroethylsulfonyl)imide (LiFPFSI), lithium (difluoromethanesulfonyl)-(trifluoromethanesulfonyl)imide (LiDFTFSI), lithium (trifluoromethanesulfonyl)(n-nonafluorobutanesulfonyl) imide (LiTNFSI). The $-SO_2-N-SO_2$ backbone in these salts affords adequate charge distribution, low dissociation, good structural flexibility, and excellent thermal, chemical, and electrochemical stability[90]. Challenges with fluorinated Li salts include their cost, high density, and the formation of a thick solid-electrolyte interphase. To overcome these obstacles, non-fluorinated salts are under investigation. Zhang *et al.*[131] investigated the fluorine-free salt anion, Lithium tricyanomethanide (LiTCM for solid-state Li-S batteries. A highly conductive thin solid-electrolyte interphase (SEI) formed on the Lithium anode, which aided in inhibiting Li metal dendritic growth and polysulfide shuttling. Figure 7d illustrates the morphological and chemical composition of the SEI layer produced in the electrolytes based on LiTCM and LiTFSI. The process of reducing LiTCM resulted in creating a polymeric C=N network and $Li_3N$, which provides the SEI layer with enhanced mechanical strength and improved conductivity for Li-ion transport. Additives to LiTCM have been shown to improve electrolyte performance. Santiago *et al.*[132] investigated a mixture of non-fluorinated (LiTCM) and fluorinated (LiDFTFSI) in addition to the LiTFSI salt. Their findings demonstrated that a small amount of additives improved interfacial characteristics, resulting in a homogeneous and compact SEI layer. The Li-S cells with the LiDFTFSI additive exhibited a cycle life of over 100, maintained a steady coulombic efficiency (CE) close to 100%, and achieved a capacity of approximately 400 mAh.$g^{-1}$ at the end of cycling at a rate of 0.1C1. The comparison of LiTFSI, LiDFTFSI and LiTCM is depicted in Figure 7a.

The capacity loss in Li-S batteries using liquid electrolytes is inevitable on account of diffusion of the LiPSs, migration of the dissolved PSs, solubility of the higher order PSs, and precipitation of the lower order PSs[133]. PEO is the most widely employed polymer in SPEs for Li-S batteries on account of its oxyethylene (EO) groups that enable the dissolution of different ionic salts. Ion transport in PEO is ascribed to cation interchain and intrachain hopping[134]. The salt/EO ratio controls the polymer-salt ionic conductivity. An excessive amount of salt impedes cation mobility. Work in progress on polymers for Li-S batteries has focused on stabilizing SEI layers, reducing the shuttle effect, lowering the polymer glass transition temperature, and raising its ionic conductivity at room temperature. PEO/LiTFSI solid polymer electrolytes have shown limited cyclability, experiencing rapid capacity fading after the first cycle. This is attributed to the properties of the SEI at the anode interface, which cannot hinder reactions between LiPSs and Li and also stop the shuttle effect. To address these issues, other salts have been investigated: LiFSI[135], LiDFTFSI[136], LiFTFSI[137], LiFPFSI[138], LiTNFSI[139] and LiTCM[131]. The structure of different Lithium salts is provided in Figure 7b. The fluorinated salts have one active functional group ($-SO_2CF_3$ and $-SO_2F$). LiFTFSI, on the other hand, has two groups and combines the attributes of LiTFSI and LiFSI. The ($-SO_2CF_3$ and $-SO_2F$ functional groups exhibit favourable Lewis base properties that can also efficiently capture and immobilize LiPSs. The coexistence of these functional groups in LiFTFSI augments the SPE stability, its ionic conductivity, and the safety of Li-S batteries[140]. The $-CF_2H$ units in LiDFTFSI were found to react with PEO and lower anion mobility, thus raising Li-ion conductivity ($2 \times 10^{-4}$ S $cm^{-1}$). Furthermore, these salts have been found to produce a stable and robust SEI on the Li anode surface and to decrease the shuttle effect. The non-fluorinated salt LiTCM generates a C=N structure and $Li_3N$ species on the Li anode surface. The C=N network can anchor LiPSs, hinder their mobility, and serve as a conduit for the efficient transport of lithium ions. The C=N network is stable at high temperatures (above 300 °C) and, thus, well-suited for Li-S batteries operating at high temperatures[141,142].

*5.2 Filler particles*

Inorganic fillers, such as oxides ($SiO_2$, $TiO_2$, $AlO_2$, $Al_2O_3$, $ZrO_2$), sulfides, and silicates have been shown to diminish the polymer crystalline fraction and augment the ionic conductivity of polymer electrolytes[73,143]. The inorganic fillers are also reported to adsorb polysulfides. Moreover, inorganic fillers can provide ion-conducting pathways on their surfaces, thereby enhancing Li-ion conductivities. In the case of ionically conducting fillers such as LLZO, in addition to surface effects, their intrinsic high conductivity can contribute to the overall SPE ion transport. The different Li ion transport pathways when the fillers are added are illustrated in Figure 7c. By creating connections with the oxygen atoms in the PEO chains, the oxide nanoparticles act similarly to a Lewis acid. This interaction reduces PEO crystallization and weakens the polymer-Li-ion bonding, increasing ion conduction[144–147]. Further modifications to these fillers can also be made to improve the efficiency of Li-S batteries. For instance, Fu *et al.*[147] introduced oxygen vacancies into $ZrO_2$ using carbon nanotubes and hydrogen-etching. Incorporating oxygen vacancies proved effective in reorganizing surface charge distribution, augmenting sulfiphilicity, and amplifying the number of active sites. This, in turn, incremented ionic and electronic transfer kinetics and minimised redox barriers between polysulfides and $ZrO_{2-x}$. Li *et al.*[148] combined nitrogen-doped carbon nanotubes with cobalt atoms, which were loaded into mesoporous $SiO_2$. The study revealed that cobalt in the modified fillers exhibits a capacity for adsorbing LiPSs and catalytic activity, thereby promoting the redox kinetics of LiPS. The amount of filler material must be optimized as excessive amounts can result in phase separation and aggregation, negatively impacting performance, as shown by Nguyen *et al.*[149]. A decrease in ionic conductivity occurred when the weight ratio in PEO/$LiClO_4$ of Al, Nb-codoped LLZO was increased from 15 wt % to 30 wt%. The ionic conductivity of CPEs has been found to be also influenced by the structural design, particle size, and chemical composition of the fillers. Recently, Wang *et al.*[32] obtained a significant improvement in the ionic conductivity of PEO-based composite solid electrolytes by incorporating transition metal carbides (NiNbCeC) embedded into mesoporous silica. The study revealed that varying the mass ratio of Cerium in the carbides significantly reduces reduction impedance from 9150 O (20% Ce) to 4000 O (5% Ce) to 1730 O (10% Ce) and enables the optimization of ionic conductivity of the composite solid electrolytes. This highlights the potential of composites comprising transition metal carbides for performance improvements of solid-state Li-S Batteries.

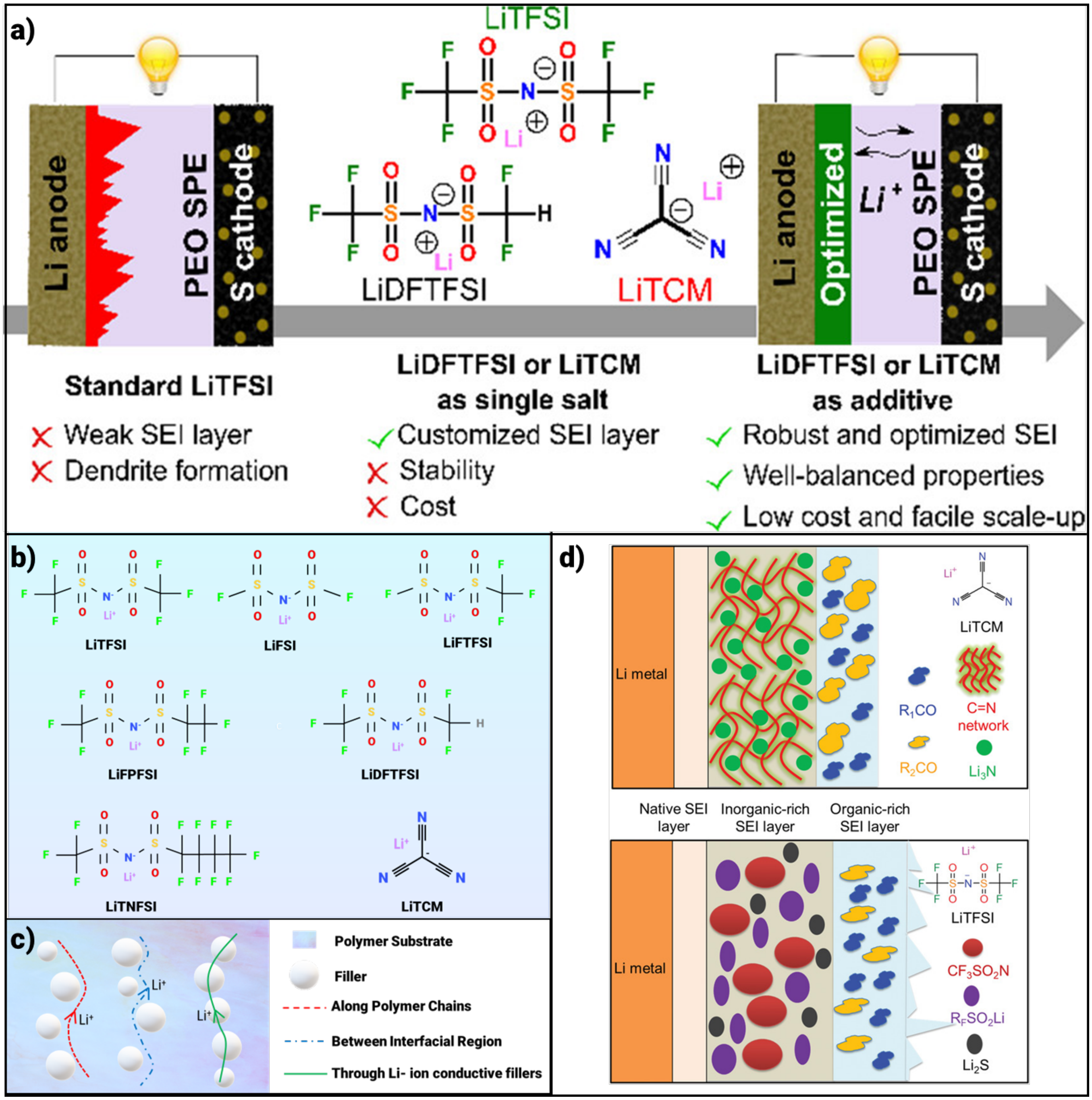

**Figure 7.** a) Attributes of various anion salts for electrolyte applications. Reproduced with permission.[132] Copyright 2021 American Chemical Society. b) Structure of different Lithium salts. c) Illustration of lithium-ion transport pathways in the presence of fillers. d) Depiction of the SEI layer formation on lithium metal anodes in electrolytes containing LiTCM and LiTFSI. Reproduced with permission.[131] Copyright 2019, John Wiley and Sons.

### *5.3 Polymer Electrolyte Design Strategies*

Another strategy to mitigate the shuttle effect involves employing polymers with a low donor number (a quantitative indicator of Lewis basicity associated with their anion salt solvating ability) for Li-S batteries, such as polymeric siloxanes. The two primary polymeric siloxane-based SPEs investigated are silsesquioxanes and polysiloxane-based electrolytes[123]. Polymeric siloxanes have low glass transition temperatures, good safety, great anti-oxidative ability, excellent heat resilience, and high compatibility with lithium metal anodes. Despite this, polysiloxanes have two inherent disadvantages: ion-insulating backbones and poor ionic dissolution, which limit their ability to transport lithium ions. To overcome this limitation, different polymer structural and compositional modifications are being pursued; this includes copolymerization through polycondensation reactions or radical polymerization reactions, grafting diverse functional side chains, cross-linking using strategies like free radical polymerization, hydrosilylation reactions, sol-gel processes, and blending with different[123]. Several polysiloxane copolymers, including comb, double-comb, cyclic, and cross-linked structures, have been used for SPEs. The ionic conductivities of polysiloxanes-based SPEs at room temperature have achieved a value of $1.2 \times 10^{-3}$ S cm$^{-1}$, while their electrochemical stability windows have been extended to 5.0 V[150,151].

*5.4 Electrode/electrolyte interfaces*

The manipulation of electrode interfaces at sulfur cathodes, separators, and lithium anodes by employing surface coatings and interlayer material modifications has shown promise for improving Li-S battery performance[135,152]. Yu *et al.*[153] formed a multi-functional, three-dimensional, thin, and dense coating of gel PVDF-HFP electrolyte on the sulfur cathode. The gel layer serves the dual purpose of restraining the volume expansion of the sulfur cathode during the discharge process and retarding the shuttle effect. The composite cathode, with 5% GPE concentration, demonstrated a consistent discharge-specific capacity of 815 mAh.g$^{-1}$ after 105 cycles, a degradation rate of 0.179% per cycle, as compared to 0.407% in the pristine CNTs/S composite cathode, indicative that the coatings effectively mitigate capacity decay of Li-S batteries. Interlayers between the separator and cathode are also found to be effective in mitigating the shuttle effect, as they can provide reaction sites for the adsorption and anchoring of polysulfides. Wang *et al*.[127] fabricated an interlayer material possessing feather-like morphology and a significantly high density of active sites through the integration of electrospinning and chemical vapour deposition (Figure 6a). The interlayer configuration involved carbon nanotubes (CNTs) on the surface of nitrogen-doped carbon fibres (N-CF), which significantly increased exposure of active Co and N-dopant adsorption-catalytic sites. The active catalytic sites facilitated the transformation of polysulfides, thereby impeding the LiPSs shuttling.

*5.5 Multifunctional polymer solid-state electrolytes*

SPEs, as already noted, provide the combined functionality of the separator and electrolyte of liquid electrolytes. In addition, the polymer matrix can be further functionalized with catalytic nanoparticles to improve the performance of Li-S batteries. Catalysts can be embedded in the polymer matrix to accelerate the reaction rates of polysulfide transformation to enhance C-rates for fast charging, by reducing the activation energy of the conversion reactions between sulfur and lithium. This results in enhanced utilization of sulfur and increased discharge capacities[60,154,155]. In another study, Zheng *et al.*[156] synthesized bidirectional catalysts (a material that can facilitate both the reduction and oxidation reactions during the charge and discharge cycles) by anchoring vanadium carbide onto N-doped carbon nanotubes (VC@NCNTs). This catalyst proved to be highly effective in adsorbing polysulfide from the electrolyte and rapidly converting it. As a result, the activation energy was significantly reduced, substantially improving both the forward and reverse redox kinetics. It was observed that including the VC catalyst on the cathode can enhance the conversion process of $Li_2S$ activation to $S_8$ to a certain degree, resulting in improved fast-charging performance.

Polymers can also host fillers to improve adhesion to the electrode. For example, the incorporation of $Li_{3.25}Ge_{0.25}P_{0.75}S_4$ (LGPS) into polymer-based electrolytes has demonstrated improved wettability of the electrolyte, thereby enhancing its contact with electrode materials. The electrochemical stability of LGPS-polymer composite electrolytes surpasses that of pure polymer electrolytes, and the improvement in the electrolyte properties resulted in enhanced electrochemical performance of the Li-S batteries[157,158].

Another strategy for developing multifunctionality in SPEs for Li-S batteries is *Defect Engineering.* Defects (vacancies, impurities, structural defects) have been employed in Li-ion batteries to increase anode capacity, accelerate redox reactions and modify the microstructure of anode and cathode materials to enhance lithiation and delithiation rates in intercalation electrodes[159]. In the case of Li-S batteries, vacancies provide excess localized electrons in vacancies to serve as effective sites for adsorption and catalysis. This enables the immobilization and conversion of different intermediates, leading to a significant enhancement in the electrochemical performance of rechargeable battery systems[160]. Introducing defects in the polymer matrix can enhance the available space, thereby promoting the mobility of Li ions and enhancing ionic conductivity. Defects have the potential to disturb the orderly arrangement of polymer crystalline regions, resulting in an augmentation of the amorphous portion. This enhanced chain mobility facilitates more efficient ion conduction[60,161]. Modifying the polymer surface with functional groups can enhance the compatibility between the polymer and the electrodes. This modification leads to a reduction in interfacial resistance and an improvement in electrochemical stability. By introducing controlled defects that function as cross-linking points, it is possible to improve the mechanical strength of the electrolyte without compromising its flexibility. Incorporating nanofillers as a defect engineering strategy can strengthen the polymer matrix, enhancing its capacity to inhibit the growth of lithium dendrites[147,162]. Introducing defects with specific functional groups makes it possible to create sites where polysulfides can be adsorbed, thereby restricting their migration. For instance, Li *et al.*[163] implemented a defect engineering strategy to introduce defects onto lightweight MOFs. The MOFs with intended defects were utilized to construct an electrocatalytic membrane. This membrane could enhance the conversion of lithium polysulfides in Li−S batteries with high sulfur loadings and a low electrolyte/sulfur (E/S) ratio. The enhanced performance was ascribed to the augmented quantity of active sites resulting from defects and the robust interactions between polysulfides and the defect sites, consequently diminishing the ill effect of shuttling. Thus, implementing defect engineering strategies in SSEs provides a pathway to potentially circumvent the limitations of Li-S batteries, currently preventing them from applications for transportation (EVs) and consumer electronics.

**Table 2**. Polymer-based Solid Electrolytes for Li-S batteries.

| Electrolyte | Electrolyte Type | Ionic Conductivity ($\mathbf{S\ cm^{-1}}$) | Thickness (µm) | Discharge ($\mathbf{mAh\ g^{-1}}$) | Cycles | Current Density | Voltage plateau discharge [V] vs Li | Ref |
|---|---|---|---|---|---|---|---|---|
| PEO/Fullerene ($C_{60}$) | CPE | $1.27 \times 10^{-4}$ (60 °C) | 100 | 1148 (1121) | 24 | 0.1 C | 2 2.3 | [104] |
| $Nb_2CT_x$ PEO/LiTFSI | CPE | $4.9 \times 10^{-4}$ (60 °C) | 100-110 | 1149 (491) | 200 | 0.5 C | 0 6 | [105] |
| HBPS/(PMMA-*block*-PPEGMA)$_{30}$/CNT/LiTFSI) | CPE | $6.58 \times 10^{-4}$ (80 °C) | - | 133 | 80 | 0.1 C | 2.3 4 | [164] |
| PEO/PGA/$Al_2O_3$/ LiTFSI/$Py_{13}$TFSI | CPE | $4.9 \times 10^{-4}$ (50 °C) | 135 | 606 (541) | 100 | 0.2 C | 1.7 2.9 | [106] |
| PTFE/LLZO/PEO | CPE | $5.03 \times 10^{-5}$ (30 °C) | 20 | 655 (568) | 100 | 0.1 C | 1.7 2.8 | [108] |
| PEO/LiTNFSI | SPE | $3.69 \times 10^{-4}$ (90 °C) | 150 | 450 (450) | 200 | 0.2 C | 2 2.4 | [165] |
| PEO/LiTFSI/LATP | CPE (sandwich type) | $4 \times 10^{-5}$ (RT) | 360 | 1793 (976) | 120 | 0.1 C | 2.05 2.3 | [89] |
| PEO/PAN/LLZO | GPE | $2.01 \times 10^{-3}$ | 100 | 942 (555) | 500 | 1 C | 1.7 2.8 | [109] |
| PEO/LiTFSI/$In_2O_3$ | CPE | $5.27 \times 10^{-4}$ (60 °C) | 90 | 695 (570) | 500 | 1 C | 1.7 2.8 | [110] |
| PEO/LiTFSI/NiO-$C_3N_4$ | SPE | $1.2 \times 10^{-4}$ (30 °C) | - | 810 (411) | 250 | 0.5 C | 1.5 3 | [111] |
| PAN/CB/VOOH-PAN/PVDF(HFP) | CPE (bilayered) | $2.81 \times 10^{-3}$ | 70 | 1391 | 500 | 0.1 C | 1.7 3.1 | [115] |
| PVdF-HFP/LiTFSI | GPE | ~$2.7 \times 10^{-3}$ | 80–100 | 1345 (400) | 20 | 0.3 mA $cm^{-2}$ | 1.7 2.8 | [114] |
| PVdF-HFP/LiTFSI/ UCPBA | CPE | $3.6 \times 10^{-4}$ (RT) | 50 | 1200 (1039) | 100 | 0.2 C | 1.0 2.8 | [42] |
| PVDF-HFP/ $\gamma$-$Al_2O_{3b}$ | GPE | $0.94 \times 10^{-3}$ | 35 | 1226.8 (841.5) | 150 | 0.1 C | 1.7 2.8 | [166] |
| Ni-Co MOF@PAN/LiTFSI | CPE | $3.06 \times 10^{-4}$ | 46 | 1560 (794) | 500 | 0.1 C | 1.7 2.9 | [116] |
| PEO-PAN-LiTFSI | SPE (cross-linked) | $1.63 \times 10^{-5}$ (RT) | 30 | 1200 (≈800) | 75 | 0.1 C | 2.0 2.4 | [34] |
| PMMA/LiTFSI | GPE | $2.44 \times 10^{-2}$ (RT) | - | 842 (264) | 200 | 0.1 C | 1.8 2.8 | [167] |
| PTMG-HDI-BHDS/LiFSI | SPE (self-healable) | $2.4 \times 10^{-4}$ (30 °C) | 200 | 602 (560) | 700 | 0.3 C | 1.0 3.0 | [120] |
| PAES-g-(PUS/2PEG90) | SPE (self-healable) | $0.659 \times 10^{-3}$ (RT) | 25 | 929.8 (917.3) | 200 | 0.2 C | 3.5 5.0 | [121] |
| BPSO/PVDF/LiTFSI/ Cellulose acetate | CPE | $4.0 \times 10^{-4}$ | 120 | 1493 (910) | 80 | 0.1 C | 1.5 3.0 | [124] |

Abbreviations used in Table 2: HBPS: Hyperbranched star polymer, PPEGMA: Poly(ethylene glycol) methyl ether methacrylate, CNT: Carbon nanotubes, LiTNFSI: lithium (trifluoromethanesulfonyl)(n-nonafluorobutanesulfonyl)imide ($Li[(CF_3SO_2)(n\text{-}C_4F_9SO_2)N]$, $In_2O_3$: Indium(III) oxide, CB: conductive carbon black, Al: Aluminum, RT: Room temperature,

## 6. Summary and Future Prospects

Li-S batteries are promising candidates for beyond Li-ion battery devices on account of their high theoretical specific capacity and energy density. Solid electrolytes can potentially advance the realization of practical Li-S batteries, as they can inhibit the shuttle effect and dendrite formation, as well as offer superior safety over liquid electrolytes. SPEs provide an integrated solution that tackles various challenges concurrently, potentially resulting in more stable, safer, and higher-performing Li-S batteries. Their capacity to furnish a physical barrier, ensure mechanical stability, enhance safety, facilitate lithium metal anodes, and deliver long-term stability renders them a notably promising strategy for the advancement of practical Li-S battery technology. Moreover, by averting the necessity for separators and liquid electrolytes, polymer electrolytes can facilitate more streamlined, compact cell architectures with the potential for enhanced energy density. Solid electrolytes can be integrated with additional strategies like cathode engineering and electrolyte additives, resulting in synergistic effects that alleviate the shuttle effect and enhance overall battery performance. A comparison of the three types of solid-state electrolytes that are being actively investigated is given in Table 1, and a summary of the performance of different polymer-based solid electrolytes is given in Table 2. However, the practical implementation of SPE is impeded by obstacles that include limited ionic conductivity, electrode interface compatibility issues, lithium dendrite formation, and slow charge/discharge rates. The following summarises specific challenges and opportunities for SPEs to enable the realization of Li-S batteries.

*6.1 Overcoming Electrode-Electrolyte Compatibility Issues*

Preventing dendrite growth and deleterious interfacial reactions between the SPE and the anode and cathode electrodes is essential to minimize ion interface resistance and ensure electrochemical and mechanical stability. Potential solutions being pursued include engineered interlayers to provide strong adhesion and mechanical strength, facilitate ion transport and hinder polysulfide diffusion. On the anode side, the primary concerns are lithium dendrite growth, inadequate interfacial stability, high interfacial resistance, and electrolyte decomposition. On the cathode side, significant concerns include inadequate interfacial contact, suboptimal ionic conductivity, volumetric alterations during cycling, polysulfide dissolution (though reduced compared to liquid electrolytes), and interfacial decomposition. Looking at the cathode, incomplete interfacial contact, poor ionic conductivity, volume changes during cycling, polysulfide dissolution (though reduced compared to liquid electrolytes), and interfacial decomposition are some of the most critical concerns. Potential solutions for these issues may encompass the construction of artificial solid-electrolyte interfaces and the optimization of electrolytes. Designing structured Li-metal anodes and forming stable SEI layers on the lithium surface may serve the purpose of protecting the anode. Developing composite cathodes embedding sulfur within porous carbon or alternative nanostructures represents a feasible solution for the cathode component. Employing metal oxides or conductive polymers on the cathode can enhance the conductivity[168–171].

*6.2 Enhancing Ion transport and energy density*

Presently, the ionic conductivity requirements for practical Li-S batteries are not met by either inorganic solid electrolytes or electrolytes based on polymers. SPEs exhibit diminished ionic conductivity at ambient temperature, while inorganic solid electrolytes, despite their superior ionic conductivity, form elevated impedance interfaces with poor adhesion and conformal bonding and inferior mechanical properties. Composite polymer electrolytes comprising polymers, anion salts and inorganic fillers have demonstrated a viable approach to enhance the performance of solid-state electrolytes. In addition, the polymer matrix can host functional particles such as catalysts and dopants that can mitigate the shuttle effect and enhance the rate of conversion of long-chain polysulfides to $LiS_2$. Polymer-based solid electrolytes exhibit high flexibility, effectively accommodate the volume expansion of sulfur cathodes, and can provide good contact and adhesion with the electrode. Further advances are expected from engineering the physicochemical properties and structure of the constituent polymer, anion salts and inorganic filler materials to solve the transport properties and thermal and mechanical properties of CPEs. Improvements in polymer properties can be derived from blending polymers that afford complementary properties such as mechanical strength, anion solubility and ionic conductivity. Work on novel filler materials rationally designed to provide defects, surface charges, and electrostatic interactions with the CPE constituent materials will significantly improve battery performance.

SPEs can also aid in reducing the total weight of the battery to increment its energy density. The electrolyte weight contribution is presently > 47.5%, although its impact on the energy output is minimal[172]. Thus, decreasing the amount of electrolyte material (lowering the electrolyte to sulfur (E/S) ratio) increases the battery energy density. Polymer thin films can be reliably fabricated with sub-micron thicknesses and pin-hole free. Thus, the diffusion length and time of ion transport between the active electrodes can be reduced by more than one order of magnitude over current liquid electrolytes. This length-scale reduction can circumvent the ionic conductivity deficiencies of SPEs in comparison to liquid electrolytes. However, said ultrathin SPEs must provide the adhesion and mechanical properties necessary for unhindered ion transport and cathode volume expansion.

## 7. Future work recommendations

We suggest that advancements and material solutions for SPEs for Li-S batteries must include the following areas for future research:

### *7.1 Polymer selection*

Polymers for Li-S batteries need to provide higher ionic conductivity than current materials, and they must be chemically and electrochemically compatible with sulfur to avoid degradation of the polymer; they must offer high solubility for the anion salts and provide the mechanical properties for thermal stability over the battery temperature operational range. It is essential that the polymer does not deteriorate or break down at high temperatures or under fast charging conditions. The polymer should not burn or emit harmful gases or materials when subjected to heat or mechanical stress. Lithium polysulfides should be soluble in the polymer while hindering the diffusion of polysulfides and enabling their fast conversion to short-chain reaction products. To this end, filler particle additives play a crucial role in CPEs. The control of lithium polysulfide solubility plays a pivotal role in enhancing the efficiency and durability of Li-S batteries. The polymer should exhibit selective cation transport and hinder the conduction of electrons to prevent short circuits. In addition, for fabrication scalability, the polymers should be low cost, and the fabrication of SPE should be scalable, reproducible, and high yield. In this review article, examples have been given on active research to achieve the aforementioned requirements, and future progress is predicated on a deeper physical understanding of the mechanisms that control transport in SPEs and on the interactions between constituent materials of SPEs.

### *7.2 Additives*

Composite polymer electrolytes have shown higher ionic conductivity and mechanical stability than their polymer base. The incorporation of nanoparticles or nanofibers into the polymer matrix improves ionic conductivity and mechanical strength. Incorporating plasticizers into the polymer matrix increases ion transport and mechanical flexibility. Plasticizers such as poly (ethylene glycol) (PEG) are employed to increase the flexibility and reduce the crystallinity of the polymer electrolyte. Thus, improving ionic conductivity by enhancing polymer chain segmental mobility. Ionic liquids can improve ionic conductivity and lower the glass transition temperature of polymer electrolytes, making them attractive candidates for Li-S batteries[173]. The mechanical strength and thermal stability of SPEs are improved by including nanoparticles such as silica, alumina, or graphene in the polymer matrix. Furthermore, these nanostructures offer additional fast ion pathways and increment the ionic conductivity of the CPEs. To immobilize polysulfide additives such as carbon nanotubes (CNTs), graphene, metal oxides, metalorganic frameworks (MOFs), and Lithium Polysulfide scavengers can be incorporated in the polymer matrix to prevent loss of active material and polysulfide diffusion. Polymers can also host catalytic particles, dopants, and defects that can play a critical role in enhancing polysulfide conversion and limiting their diffusion, thereby enhancing the rate of chemical reactions and hindering the shuttle effect. Further progress on the design of custom nanoparticles to improve the functionality of SPEs for Li-S batteries also requires an essential understanding of the relationship between filler surface properties (charges, chemistry, morphology), their dielectric and electrostatic interactions with the polymers and anion salts.

*7.3 Sulfur compatibility and interface engineering*

SPEs must be chemically stable in the presence of sulfur and its discharge products, such as lithium sulfides, as this will cause capacity fading and low cycle life with continued cycling if the electrolyte degrades or reacts with these species. The contact between the sulfur electrode and the solid polymer electrolyte needs to be stable and the formation of a robust interphase (SEI) on the sulfur electrode is crucial for avoiding undesired side reactions and improving overall Li-S battery performance. The SPE mechanical properties must accommodate the electrode volume expansion/contraction associated with charge/discharge cycling. Certain electrolyte additions, such as lithium salts and lithium-conductive fillers, can improve SPE compatibility with sulfur. These additions can aid in the formation of stable SEIs on the sulfur electrode.

As discussed in this review, filler particles, additives, and defects also offer opportunity areas for the rational design of interfaces to optimize ion transport, fast charging and thermal and structural stability and reliability of Li-S batteries.

**Author Contributions:** Conceptualization, S.C., E.E.M. and P.B.T.; writing-original draft preparation, P.B.T.; writing-review and editing, P.B.T. and E.E.M. All authors have read and agreed to the published version of the manuscript.

**Acknowledgements**: P.B.T. wants to acknowledge the Science and Engineering Research Board, India, for providing the Overseas Visiting Doctoral Fellowship (OVDF) and supporting this work at Purdue University, USA

**Conflicts of Interest:** The authors declare no conflicts of interest.